\documentclass{article}
\usepackage{spconf,amsmath,graphicx}
\usepackage{multirow}
\usepackage{hyperref}
\title{DC-WCNN: A deep cascade of wavelet based convolutional neural networks for MR Image Reconstruction} 
\name{
\begin{tabular}{@{}c@{}}
Sriprabha Ramanarayanan$^{1,2\star}$ \thanks{$^{\star}$ Contributed equally} \qquad Balamurali Murugesan$^{1,2\star}$ \qquad Keerthi Ram$^{2}$  \\ Mohanasankar Sivaprakasam$^{1,2}$
\end{tabular}
}

\address{$^{1}$ Indian Institute of Technology Madras (IITM), India \\ $^{2}$ Healthcare Technology Innovation Centre (HTIC), IITM, India}

\begin{document}
%\ninept
%
\maketitle
\begin{abstract}
Several variants of Convolutional Neural Networks (CNN)
have been developed for Magnetic Resonance (MR) image
reconstruction. Among them, U-Net has shown to be the
baseline architecture for MR image  reconstruction. However, sub-sampling is performed by
its pooling layers, causing information loss which in turn
leads to blur and missing fine details in the reconstructed
image. We propose a modification to the U-Net architecture to recover fine structures. The proposed network is a wavelet packet transform based
encoder-decoder CNN with residual learning called WCNN. The proposed WCNN has discrete wavelet transform instead of pooling and inverse wavelet transform instead of unpooling layers and residual connections.  We also propose a deep cascaded framework (DC-WCNN) which consists of cascades of WCNN and k-space data fidelity units to achieve high quality MR reconstruction. Experimental results show that
WCNN and DC-WCNN give promising results in terms of evaluation metrics and better recovery of fine details as compared to other methods.
\end{abstract}


\section{Introduction}
Magnetic Resonance Imaging (MRI), an anatomical non-invasive imaging technique is well known for providing high-resolution images with excellent soft tissue contrast. The key challenge however is to reduce its long scan times to ease patient discomfort. Consequently, sub-Nyquist sampling is adopted in k-space to accelerate data acquisition from the modality and this naive reconstruction from sampled k-space results in images with aliasing artifacts. Approaches based on Compressed Sensing (CS-MRI) aim to solve the de-aliasing problem by enforcing image sparsity and incoherent sampling in k-space \cite{Hollingsworth_2015}.  
\begin{figure}
  \centering
  \includegraphics[width=0.9\linewidth]{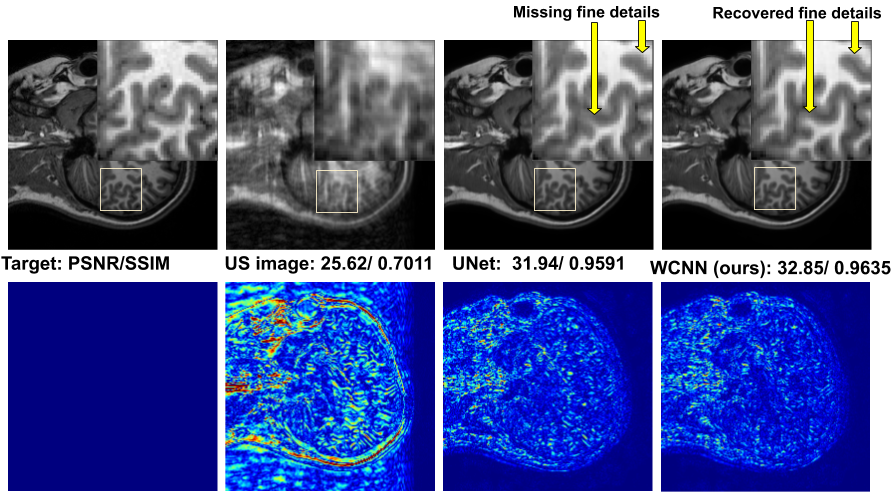}

\caption{Top row (left to right): Human brain target image with highlighted region in yellow box, shows fine details, US image (with 5x acceleration factor) has artifacts, UNet output suffers from blur, WCNN (ours) shows better recovery of fine structures. Bottom row: reconstruction errors with respect to target show that WCNN gives minimal error as compared to U-Net.}
  \label{fig:front}
\end{figure}

Recently, deep learning methods for CS-MRI, based on Convolutional Neural Networks (CNN) that learn an end-to-end mapping between the undersampled (US) and fully-sampled (FS) image in a data driven manner has gained focus \cite{mri_survey}. Wang \textit{et al.} used a simple CNN to learn the mapping between US and FS images \cite{wang}. Subsequently, an encoder-decoder CNN based architecture, U-Net \cite{unet}, has shown promising results in many image-to-image problems.
% Later, an encoder-decoder CNN based architecture U-Net has shown promising results in many image-to-image problems because of the following reasons: 1) ability of the encoder to provide features in a low-dimension latent representation 2) ability of decoder to preserve fine details through skip connections from the encoder \cite{unet}. 
Lee \textit{et al.} used U-Net to learn the residual and showed that their method gives better reconstruction with faster training convergence \cite{deep_residual}. Hyun \textit{et al.} demonstrated the effectiveness of U-Net followed by k-space correction to provide consistency in the reconstructed k-space data \cite{data_consistency}. Inspired by Dictionary Learning MRI (DLMRI), Schlemper \textit{et al.} proposed a  deep cascaded CNN (DC-CNN) which is a cascade of several CNNs and Data fidelity (DF) units \cite{cascade}. Recent work by Sun \textit{et al.} replaced CNN with U-Net in the deep cascaded network (referred from hereon as DC-UNet). DC-UNet has shown improved reconstruction for higher acceleration factors \cite{segnetmri}. 

% The above mentioned approaches have warranted the choice of U-Net for MRI reconstruction. The success of U-Net is backed by the low-dimensional manifold constraint learnt by its latent space from the training data. The learned manifold constraint acts as an efficient non-linear compressed sensing, mapping an under-sampled MRI image to a fully sample MRI image. The second important motivation to choose UNet is its receptive field. The hierarchical levels of scale due to convolution and pooling layers of Unet helps to enhance the receptive field exponentially  \cite{deepConvFramelets}. Translation invariance in its pooling layers is the third factor which has shown to be very beneficial for high level vision tasks like image classification \cite{pooling}. However, the sub-sampling step in  the pooling results in information loss and in turn degrades the signal information, an undesirable factor for low-level vision tasks like image reconstruction. A fully convolutional network (FCN) without pooling/unpooling layers in U-net could still address this limitation but cannot provide the desired receptive field \cite{deepConvFramelets}. Dilated convolution provide good receptive field but produce unwanted checkerboard artifacts and inconsistency at the edge regions of the features owing to its sparse sampling property.  
Although originally created for segmentation tasks, the use of the U-Net in the above papers indicate that it could be a good baseline architecture for US to FS reconstruction. This could be attributed to the following reasons: 1) The latent space representation learned from the training data acts as highly nonlinear compressed sensing to provide a mapping between US and FS images and 2) The hierarchical scale levels due to convolution and pooling layers provide good receptive field. However, pooling layers in U-Net causes information loss and in turn degrades the signal information, especially fine structural details, an undesirable factor for image reconstruction. A fully convolutional network (FCN) without pooling layers in U-Net could still address this limitation but cannot provide the desired receptive field \cite{deepconvframelets}. Dilated convolution can be used in place of normal convolution in FCN to provide the necessary receptive field. But, dilated convolution introduces unwanted checkerboard artifacts and shows inconsistency at the edge regions of the features owing to its sparse sampling property \cite{DilatedConv}.

% From these observations, the information loss in the feature maps is apparent. Though 

% The signal degradation due to loss of information has been addressed by adopting several network design strategies (some of them include residual connection and channel attention) have been developed to improve the representational power of the network, yet do not address the limitation in terms of preserving signal details. To address this issue inherent in pooling layers, we propose a suitable transformation from the classical signal processing theory to augment the pooling functionality. 
% % The basis vectors of the proposed transformation interact with the signal in a global context to capture frequency and location information.
% We have chosen multi-level Haar wavelet packet transform (WPT) for its high relevance in deep learning ([],[],[]) in terms of 1) Frequency and localization characteristics that help in efficient representation of structural and textural details 2) Orthonormality, energy compaction and multi-resolution formulation that provide lossless partitioning and recovery, signal energy preservation and multi-scale representation of feature maps. 3) Sparsity induced in feature maps at every resolution level in the network that reduces the overall computational complexity []. 
% Related work in wavelet domain neural network architectures include [], [], [].
% (WavResNet, DWSR, Wavelet SRNet, Subband Regularized CNN, Deep Convolutional Framelets).

In our work, we propose to replace the pooling layers with a transformation inspired from classical signal processing theory to circumvent the information loss. We choose multi-level Haar wavelet packet transform (WPT) as it can provide hierarchical scale levels without information loss. The salient features of WPT which makes it highly appropriate for deep learning include: 1) Frequency and localization characteristics which helps in efficient representation of structural and textural details 2) Lossless partitioning of feature maps into orthogonal subbands 
% and energy compaction 
at multiple scales 
% Orthonormality, energy compaction and multi-resolution formulation provides lossless partitioning and recovery, signal energy preservation and multi-scale representation of feature maps 
and 3) Sparsity induced in feature maps at every resolution level which reduces the overall computational complexity. Fig. \ref{fig:front} highlights above mentioned merits of using wavelet transforms in U-Net for recovery of fine details. Combining these advantages, we summarize our contributions as follows:
% \subsection{Contributions}
\begin{itemize}
\item We propose a wavelet based convolutional encoder-decoder neural network WCNN, for MR image reconstruction,  with better signal representation,
% and receptive field, 
inspired by the work in \cite{MWCNN} for vision tasks. The proposed WCNN has residual connections, wavelet decomposition and re-composition operations in place of pooling and unpooling layers respectively. 
\item We also propose a deep cascaded architecture called DC-WCNN, by cascading a series of WCNN and data fidelity (DF) units for the MRI reconstruction problem.  

    % \item Inspired by the work in [] for vision tasks, we propose a WPT based convolutional encoder-decoder neural network, WCNN, with layer-wise residual connections and wavelet decomposition and re-composition operations in place of pooling and unpooling layers respectively for lossless extraction of feature maps, preserving high resolution details with better receptive field.
    % \item We propose an overall deep cascaded network (DeepCascadeWCNN) with cascades of WCNN block and data consistency unit for efficient end-to-end optimization of the MR reconstruction problem. 
    % % To the best of our knowledge we are the first to introduce a wavelet based Deep cascade CNN framework for MRI Reconstruction. 
    \item Performance analysis of WCNN and DC-WCNN with Kirby21 brain dataset shows promising results for the given 5x undersampling mask and outperforms other compared methods.
\end{itemize}

\section{Methodology}

\begin{figure*}
  \centering
  \includegraphics[width=0.90\linewidth]{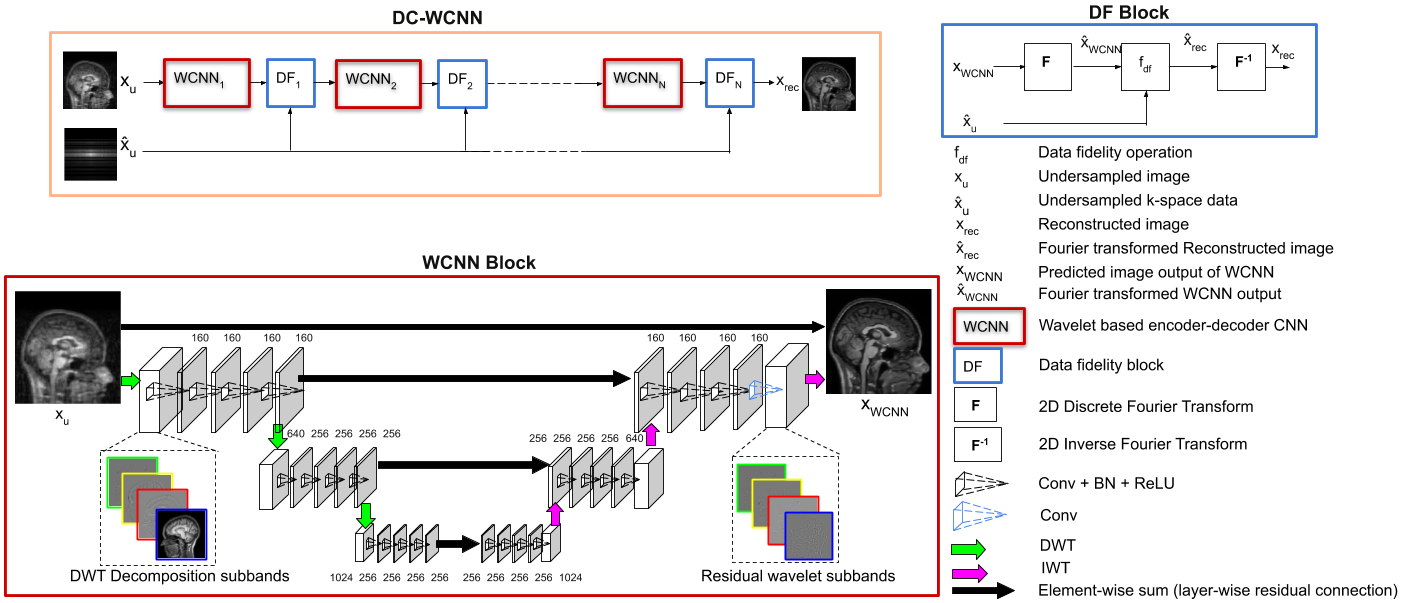}
  \caption{DC-WCNN: Proposed architecture for MRI Reconstruction}
  \label{fig:architecture}
\end{figure*}

\subsection{Problem Formulation}
% The problem formulation starts by first defining the DL-MRI problem and inheriting the same for our DC-WCNN formulation. Given an image, $x\in C^{P}$, $x_{ij}\in C^{n}$ is
% the vector representation of the square 2D image patch of size $\sqrt n \times \sqrt n$ pixels, indexed by the location of its top-left corner $(i,j)$ in the image. Let $D \in C ^{n \times K}$ represent the image patch based dictionary with K columns.
% DL-MRI aims to solve the following optimization problem.
% \begin{equation}
% \underset{D,\Gamma}{\operatorname{argmin}} \sum_{ij}\|R_{ij}x - D\alpha_{ij} ||_{2}^2 + \beta||\alpha_{ij} ||_{0}
% \end{equation}
% The matrix $R_{ij}\in {C}^{N}$ x represents the operator that extracts the patch $x_{ij}$ from x as $x_{ij}=R_{ij}x$ . The quasi norm $l_{0}$ is used to
% encode the sparsity of the patch representation. $\Gamma$ is used to denote the set of
% sparse representations of all patches. This learning formulation
% minimizes the total fitting error of all image patches with respect
% to the dictionary, subject to sparsity constraints.
% The reconstruction of FS image from US image using the proposed WCNN is formulated as,
% \begin{equation}
% \underset{x_{wcnn},\theta}{\operatorname{argmin}} ||x - x_{wcnn} ||_{2}^2 + \alpha||F_{u}x - y ||^2_{2}
% \end{equation}
Let $x\in C^{N}$ be the desired image to be reconstructed from undersampled k-space measurements $y\in C^{M}$, $M<<N$, such that $y = F_{u}x$, where $F_{u}$ is the undersampled Fourier encoding matrix. 
% The linear inversion $x_{u} = F_{u}^{H}y$, called the zero-filled reconstruction is ill-posed and generates an aliased image due to sub-Nyquist sampling.
The linear inversion $x_{u} = F_{u}^{H}y$ is called  zero-filled reconstruction. Reconstructing $x$ from $y$ is an ill-posed problem due to sub-Nyquist sampling.
The proposed method using WCNN can be formulated as the optimization problem:
% \widetilde{x} =
\begin{align}
 \underset{x,\theta}{\operatorname{argmin}} ||x - f_{wcnn}(x_{u}\lvert\theta) ||_{2}^2 
+ \alpha||F_{u}x - y ||^2_{2}
\end{align}

Let $x_{wcnn} = f_{wcnn}(x_{u}\lvert\theta)$, where $f_{wcnn}$ is the forward mapping of WCNN parameterized by $\theta$. Here $\alpha$ is a weight factor based on the noise of the acquired data y. The above equation enforces $x$ to be approximated by the reconstruction of WCNN in image domain without any prior information about the acquired data in k-space. 
% The second term represents data fidelity in k-space.

We use data fidelity (DF) unit in k-space domain after WCNN unit to ensure that the WCNN reconstruction is consistent with the acquired k-space measurements. The data fidelity operation $f_{df}$ can be expressed as,
  \begin{equation}
    \hat{x}_{rec}=
    \begin{cases}
      \hat{x}_{wcnn}(k)  & \ k\notin\Omega \\
      \frac{\hat{x}_{wcnn}(k) + \lambda \hat{x}_{u}(k)}{1+\lambda} & k\in\Omega \\
    \end{cases}
  \end{equation}
  
Here, $\hat{x}_{wcnn} = F_{f} x_{wcnn}$, $\hat{x}_{u}= F_{f} x_{u}$, $\Omega$ is the index set of known k-space data, $F_{f}$ is the Fourier encoding matrix,  and $\hat{x}_{rec}$ is the corrected k-space and $\lambda\to\infty$. The reconstructed image is obtained by inverse Fourier encoding of $\hat{x}_{rec}$, i.e. ${x}_{rec} = F_{f}^{H}\hat{x}_{rec}$. 
The proposed cascaded architecture, DC-WCNN, is a series of $N_{c}$ such WCNN and DF units which can be formulated as,
\begin{align}
x_{n} &= WCNN_{n}(x_{n-1}^{df}) + x_{n-1}^{df} \\ 
x_{n}^{df} &= DF_{n}(x_{n})
\end{align}

Here $WCNN_{n}$ and $DF_{n}$ denote the $n^{th}$ WCNN reconstruction block and DF unit respectively, $n=1,2..N_{c}$, $x_{0}^{df} = x_{u}$ and $x_{rec}=x_{N_{c}}^{df}$ is the output of the last DF unit.

The 2D Discrete Wavelet Transform (DWT) and inverse wavelet transform (IWT) layers of WCNN are given by,
\begin{align}
(X_{1}, X_{2},..., X_{K}) = DWT(X_{in}) \\ 
X = IWT(X_{1}, X_{2},..., X_{K})
\end{align}
Here, $X_{in}$ be the input image or an intermediate feature map in the network, K is the number of subbands (K $=$ 4 in our case, with approximation, horizontal, vertical and diagonal subbands), $X_{k}, k = 1,2, ... K$ denote the coefficients of the $k^{th}$ subband and X is output of wavelet recomposition.

% The proposed WPT layer in the WCNN is formulated as follows. Let $X_{in}$ be the input image or an intermediate feature map at any resolution level i in the network.The wavelet decomposition into subbands using 2D Discrete Wavelet Transform (DWT) can formulated as
% inverse wavelet transform (IWT)
% \begin{equation}
% (X_{i}^{1}, X_{i}^{2},...X_{i}^{K}) = DWT(X_{in}, K, i)
% \end{equation}
% \normalsize
% The inverse wavelet transform (IWT) at level i, is given by,
% \begin{equation}
% X = IWT(X_{i}^{1}, X_{i}^{2},...X_{i}^{K})
% \end{equation}

% Let $x_{f} \in {C}^{N}$ be the fully sampled complex image with dimensions $\sqrt N \times \sqrt N$ arranged in column-wise manner. $x_{f}$ is obtained from fully sampled k-space measurements ($y_{f} \in {C}^{N}$) through a fully sampled encoding matrix $F_{f}$ using the relation $y_{f}=F_{f}x_{f}$. During undersampling, a subset of k-space measurements ($y_{u} \in {C}^{M}$) say ($M<<N$) only are made. This corresponds to an undersampled image $x_{u}$ by the relation $x_{u} = F_{u}^{-1}y_{u}$. $x_{u}$ will be aliased due to sub-Nyquist sampling. Reconstructing $x_{f}$ directly from $y_{u}$ is ill-posed and direct inversion is not possible due to under-determined nature of system of equations. In our approach, we use deep learning network to learn the mapping between $x_{u}$ and $x_{f}$. The neural network thus learns to minimize the error between predicted fully sampled image ($\hat{x}_{f}$) and the ground truth ($x_{f}$).
\vspace{-3mm}
\subsection{Background of the proposed architecture} 
 
% A similar approach has been proposed in \cite{deepconvframelets}, wherein only the low pass subband coefficients are passed to convolution layers, while other subbands are bypassed from convolution layers and concatenated directly to the expansion layers. 
Several studies have been done to bring wavelet transforms into CNNs (\cite{deepconvframelets}, \cite{DWSR}) . The proposed WCNN is a variant of U-Net with contraction and expansion paths wherein DWT and IWT are performed in place of pooling and unpooling layers respectively.
The DWT subbands are stacked and passed to the convolutional layers. 
% This enables the subbands to be learnt together enhancing spatial contextual details. The non-linear dependencies between the subbands are also modeled. 
In this way, all the subbands are jointly learnt along with the inter-dependencies between them  thereby enhancing spatial context. 
% The four Haar basis filters are,
% \scriptscriptstyle
% \[
% \phi_{LL} = 
% \begin{bmatrix}
% \scriptscriptstyle
%     1  &  \scriptscriptstyle 1      \\
%     \scriptscriptstyle 1  &  \scriptscriptstyle 1      
% \end{bmatrix}
% \phi_{LH} = 
% \begin{bmatrix}
%     \scriptscriptstyle -1  &  \scriptscriptstyle -1      \\
%     \scriptscriptstyle 1  &  \scriptscriptstyle 1      
% \end{bmatrix}
% \phi_{HL} = 
% \begin{bmatrix}
%     \scriptscriptstyle -1  &  \scriptscriptstyle 1      \\
%     \scriptscriptstyle -1  &  \scriptscriptstyle 1      
% \end{bmatrix}
% \phi_{HH} = 
% \begin{bmatrix}
%     \scriptscriptstyle 1  &  \scriptscriptstyle -1      \\
%     \scriptscriptstyle -1  &  \scriptscriptstyle 1      
% \end{bmatrix}
% \] 
% \normaltext
% The average (LL) subband is similar to the mean pooling operation of the standard U-net, which is a smoothing operation. The concatenation of three detail subbands suppress the blurring emphasized by smoothing and thereby help the convolutional layers to learn kernels that maximize the overall performance. The multi-level WPT is shown in Figure.
We have used residual connections wherein feature maps from contracting paths are element-wise added to the respective expanding path feature maps instead of concatenations. The residual learning strategy simplifies the optimization process and minimizes degradation of original signal information. 

The WCNN as a standalone architecture unit could de-alias the US MRI image in a single step. The DC-CNN proposed by Schlemper et al. showed that a deep cascade of CNNs is similar to unfolding the optimization process of CSMRI and improves the performance of MRI reconstruction. 
We therefore embed WCNN as the reconstruction block in deep cascaded mode with DF units interleaved to form the DC-WCNN architecture.
\vspace{-3mm}
\subsection{Proposed architecture}
The proposed DC-WCNN has cascades of WCNN and DF units (Fig.2). The WCNN consists of three WPT levels (i.e. three subsampling layers in the network). The subband images obtained from each level of DWT is fed to a 4-layer fully convolutional (FC) block in the contraction path. In the expansion path, IWT is performed after every FC block. The convolution layers have 3x3 convolution filters followed by batch normalization (BN) and rectified linear unit (ReLU) operations. In the last layer, only convolution operation is performed without BN and ReLU, to predict the residual subbands. 
In the DC-WCNN, the DF layers are inserted between WCNN units to correct the accumulation of possible distortions in the predicted k-space data over the cascaded blocks.  
% Unet archtecture with enhanced receptive field
% Residual connection
% Talk about the deep cascade with wavelet unet architecture. %Deep cascade blocks with wavlet unet as convolution blocks followed by data consistency layer. 
%Each block has a connection from the input as it enough to learn the residual 
% \subsection{Loss function}
We have used L2 loss function for both standalone and cascaded modes. Given the training data D consisting of a number of US and FS images as  input-target pair $(x_{u}, x_{t})$ as given by,
\begin{align}
L(\theta) = \sum_{(x_{u}, x_{t})\in D}\|x_{t} - x_{pred} ||_{2}^2 
\end{align}
Here $x_{pred} = x_{wcnn}$ in the standalone mode and $x_{pred} = x_{rec}$ in the deep cascade mode.
% The training dataset consisting of N US input image-FS target image pairs,  $(x_{u}, x_{t})$ and $x_{cnn}$ denote the reconstructed image of DeepcascadeWCNN, the loss function can be defined as, 
% Loss function l1 loss. 
% If we have time: Extra terms like ssim or vgg loss. We can say our network + ssim loss performs better than our network without ssim loss. 
\vspace{-3mm}
\section{Experiments and Results}
\vspace{-2mm}
\subsection{Dataset description and Evaluation metrics}
We have conducted experiments on the publicly available Kirby21 dataset \cite{Kirby} with human brain data. The dataset consists of 5460 slices of size 256x256 taken from 42 T1-weighted MPRAGE volumes out of which, 3770 slices from 29 volumes are used for training and 1690 slices from 13 volumes for validation. A fixed cartesian undersampling mask with ten lowest spatial frequencies and remaining following a zero-mean Gaussian distribution is chosen. The acceleration factor for the mask is 5x (20\%) to retrospectively generate undersampled MRI images for training and testing. Peak Signal-to-Noise Ratio (PSNR), Structural Similarity Index (SSIM), High Frequency Error Norm (HFEN) and Normalised Mean Square Error (NMSE) metrics are used to evaluate the reconstruction quality. Wilcoxon signed-rank test with an alpha of 0.05 is used to assess statistical significance.
\begin{table*}[]
\scriptsize
\centering
\caption{PSNR, SSIM, NMSE and HFEN results for 5x undersampling}
\label{tab:my-table}
\begin{tabular}{|c|c|l|c|c|c|c|}
\hline
\multirow{2}{*}{} & No. of cascades & \multicolumn{1}{c|}{Model} & NMSE & PSNR & SSIM & HFEN \\ \cline{2-7} 
 & - & US image & 0.07014 +/- 0.01 & 25.95 +/- 1.29 & 0.6056 +/- 0.05 & 0.7571 +/- 0.01 \\ \hline
\multirow{3}{*}{Standalone mode} & \multirow{3}{*}{-} & CNN \cite{wang} & 0.02775 +/- 0.00 & 29.97 +/- 1.33 & 0.8897 +/- 0.02 & 0.5311 +/- 0.07 \\ \cline{3-7} 
 &  & U-Net \cite{unet} & 0.01929 +/- 0.01 & 31.57 +/- 1.17 & 0.9302 +/- 0.01 & 0.4583 +/- 0.06 \\ \cline{3-7} 
  &  & U-NetMean & 0.01955 +/- 0.00 & 31.51 +/- 1.21 & 0.9303 +/- 0.01 & 0.4574 +/- 0.06 \\ \cline{3-7}
 &  & WCNN (Ours) & \textbf{0.01466 +/- 0.00} & \textbf{32.75 +/- 1.55} & \textbf{0.9361 +/- 0.02} & \textbf{0.4228 +/- 0.06} \\ \hline
\multirow{9}{*}{Deep cascade mode} & \multirow{3}{*}{1} & DC-CNN \cite{cascade} & 0.0179 +/- 0.00 & 31.88 +/- 1.50 & 0.8784 +/- 0.02 & 0.4776 +/- 0.07 \\ \cline{3-7} 
 &  & DC-UNet \cite{segnetmri} & 0.01181 +/- 0.00 & 33.69 +/- 1.56 & 0.9294 +/- 0.02 & 0.4033 +/- 0.06 \\ \cline{3-7} 
 &  & DC-WCNN (Ours) & \textbf{0.01151 +/- 0.00} & \textbf{33.8 +/- 1.61} & \textbf{0.9308 +/- 0.02} & \textbf{0.3879 +/- 0.06} \\ \cline{2-7} 
 & \multirow{3}{*}{2} & DC-CNN \cite{cascade} & 0.01249 +/- 0.00 & 33.45 +/- 1.69 & 0.9385 +/- 0.02 & 0.4229 +/- 0.07 \\ \cline{3-7} 
 &  & DC-UNet \cite{segnetmri} & 0.00884 +/- 0.00 & 34.95 +/- 1.70 & 0.9587 +/- 0.01 & 0.357 +/- 0.064 \\ \cline{3-7} 
 &  & DC-WCNN (Ours) & \textbf{0.00811 +/- 0.00} & \textbf{35.33 +/- 1.83} & \textbf{0.9614 +/- 0.01} & \textbf{0.3458 +/- 0.066} \\ \cline{2-7} 
 & \multirow{3}{*}{3} & DC-CNN \cite{cascade} & 0.01029 +/- 0.00 & 34.3 +/- 1.82 & 0.9526 +/- 0.02 & 0.3912 +/- 0.07 \\ \cline{3-7} 
 &  & DC-UNet \cite{segnetmri} & 0.0077 +/- 0.00 & 35.56 +/- 1.77 & 0.9652 +/- 0.01 & 0.3378 +/- 0.07 \\ \cline{3-7} 
 &  & DC-WCNN (Ours) & \textbf{0.00682 +/- 0.00} & \textbf{36.09 +/- 1.96} & \textbf{0.9685 +/- 0.01} & \textbf{0.3236 +/- 0.06} \\ \hline
\end{tabular}
\end{table*}
\vspace{-2mm}
\subsection{Implementation details} We have trained WCNN as standalone and within the deep cascade framework (DC-WCNN). All the standalone models are trained for 150 epochs on  Nvidia GTX-1070 GPUs. In the DC-WCNN, the weights of each WCNN are initialized with those of the standalone WCNN and subsequently weights are fine tuned. The same strategy is followed for other models in deep cascade mode. The models are implemented in PyTorch and code is publicly available  \footnote{\href{https://github.com/sriprabhar/DC-WCNN}{https://github.com/sriprabhar/DC-WCNN}}. Adam optimizer is used with a learning rate of $0.001$. 

\begin{figure}
  \centering
  \includegraphics[width=0.95\linewidth]{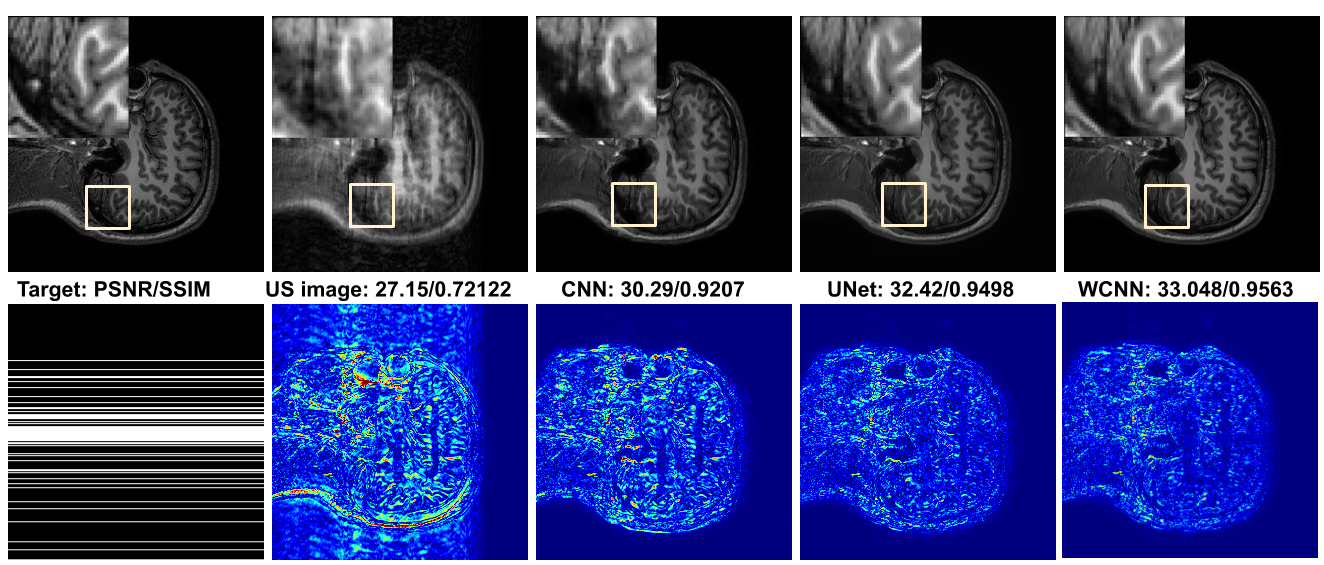}
  \caption{Standalone case (Top row from left to right): Target, US image, CNN \cite{wang}, UNetMean \cite{unet} and WCNN (ours). Bottom row: 5x sampling mask, reconstruction errors}
  \label{fig:results1}
\end{figure}
\vspace{-3mm}
\subsection{Results and discussion}
% We compare our test results with other networks in the standalone and deep cascade scenarios (refer Table 1). For the standalone case, we compare WCNN with the vanilla CNN block used in \cite{wang} and U-Net \cite{unet}. 
% % (with max pooling) and U-Net with mean pooling layers (named as U-NetMean in Table 1). 
% Our WCNN model outperforms all other methods with best values for PSNR, SSIM, NMSE and HFEN. The scores were found to be statistically significant (p$<$0.05).
% For the deep cascade case, we compare our DC-WCNN with DC-CNN \cite{wang} and DC-UNet \cite{segnetmri}. The number of cascaded blocks is varied from 1 to 3. In general, as the number of cascaded deep networks and DF units increase, improvement in PSNR and SSIM is noticed. This effect closely resembles the iterative optimization of the conventional CSMRI reconstruction. When WCNN is used as the reconstruction block of the deep cascaded structure, we achieve the best values of PSNR, SSIM, NMSE and HFEN among the other techniques. 
% % This shows that our DC-WCNN is very much appropriate for the MR reconstruction problem.
% We visually compare WCNN with vanilla CNN and U-Net in standalone mode (Fig. 3). For deep cascade mode, we compare DC-WCNN with DC-CNN and DC-UNet (Fig. 4). In standalone mode, we show that WPT based sub-sampling recovers fine details better than CNN and U-Net. In deep cascaded mode, DC-CNN and DC-UNet produce smudged structures whereas our method recovers structures much closer to the target.
\begin{figure}
  \centering
  \includegraphics[width=0.95\linewidth]{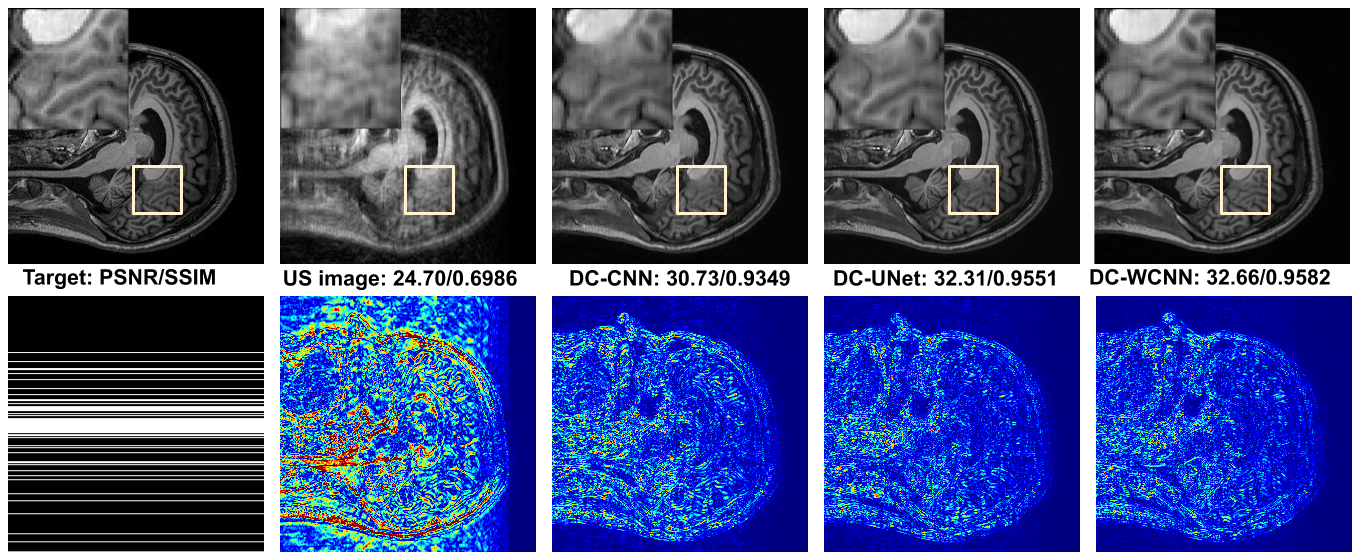}
  \caption{Deep cascade case (Top row from left to right): Target, US image, DC-CNN \cite{cascade}, DC-UNet \cite{segnetmri} and DC-WCNN (ours). Bottom row: 5x sampling mask, reconstruction errors}
  \label{fig:results2}
\end{figure}
% \subsection{Implementation details}
% Used pytorch; weights used those kind of things. Something which has to be conveyed to the reviewer. Relevant to parameters. 
We conduct two sets of experiments, one in standalone mode and the other deep cascade mode. 
% \begin{itemize}

\textbf{Standalone mode}: We compare WCNN with vanilla CNN \cite{wang} and U-Net \cite{unet} (which commonly has max pooling layers) and U-Net with mean pooling instead of max pooling layers (we refer to it as U-NetMean in Table 1). Table 1 standalone mode shows our WCNN model outperforms all other methods with best values for PSNR, SSIM, NMSE and HFEN. We note that WCNN has only 3 subsampling levels and performs better than the two U-Net architectures which have four pooling (sub-sampling) layers each. Visual comparison in Fig. 3  shows that that WPT based sub-sampling in WCNN recovers fine details better than CNN and U-NetMean (U-Net not shown as it performs same as U-NetMean). The approximation subband obtained from DWT is similar to the average pooling performed in U-NetMean, which is basically a smoothing operation. By including the detail subbands into the convolution layers, emphasis on the blurring induced by smoothing is reduced, thereby helping convolutional layers to learn kernels that maximize the overall performance of WCNN.

\textbf{Deep cascade mode}: In this mode, we compare DC-WCNN with DC-CNN \cite{cascade} and DC-UNet \cite{segnetmri}. From Table 1 deep cascade mode we make two observations. Firstly, the deep cascade mode boosts up the quantitative metrics as compared to standalone mode. This shows that deep cascaded architectures with data fidelity units provide better reconstruction. Secondly, our DC-WCNN model inherits the benefits of WCNN and outperforms other methods quantitatively. Visual comparison in Fig. 4 shows that DC-CNN (with two cascades of WCNN and DF) and DC-UNet produce smudged structures whereas our method recovers structures much closer to the target.

In both the modes, the metrics  are found to be statistically significant (p$<$0.05). We also note that increasing the number of WPT levels increases the computational cost and hence we have chosen three levels .
% Conv framelets - approx coeff is equivalent to mean pooling
% dc - adding to the benefits of WCNN, data consistency and cascaded structures improve the quality - results adhere to literature of deep cascaded framework.
% resons why WCNN gives better = becuse of all subbands,

% 1. Mean pooling equivalent 
\vspace{-3mm}
\section{Conclusion}
\vspace{-2mm}
We propose a wavelet-based CNN (WCNN) and a deep cascaded architecture called DC-WCNN for recovering fine details in MR image reconstruction. The WCNN is a variant of U-Net with DWT in place of pooling and IWT in place of unpooling. The DC-WCNN architecture is a cascade of WCNN and DF units. Experimental results show that WCNN and DC-WCNN outperform other compared methods for 5x acceleration factor.
% The proposed methods have shown improvements in reconstructed image quality due to its lossless sub-sampling and recovery of fine details.
\newpage
\bibliographystyle{IEEEbib}
\bibliography{isbi}

\end{document}